\documentclass[aps,preprint,superscriptaddress,showpacs]{revtex4}

\usepackage{graphicx}
\def\Vec#1{\mbox{\boldmath $#1$}}
\usepackage{enumerate}
\begin{document}

\title{Theory of the inelastic impact of elastic materials}

\author{Hisao Hayakawa}
\email[E-mail: ]{hisao@yuragi.jinkan.kyoto-u.ac.jp}
\affiliation{Department of Physics, Yoshida-south campus, 
Kyoto University, Kyoto 606-8501, Japan
 }

\author{Hiroto Kuninaka}
\affiliation{Graduate School of Human and Environmental Studies, 
Kyoto University,  Kyoto 606-8501, Japan}

\date{\today}

\begin{abstract}
We have reviewed recent developments of the theory of the impact for
 macroscopic elastic materials.
This review includes
(i)  standard theories for the normal impact and the oblique impact, 
(ii) some typical approaches to simulate impact problems,
(iii) and an example of our simulation to clarify
 the mechanism of  anomalous restitution coefficient
 in an oblique impact in which the restitution coefficient exceeds unity.

\end{abstract}

\maketitle

\section{Introduction}

Impacts are common in nature. Besides microscopic impacts for atoms and
molecules, there are plenty of examples of impacts for macroscopic
materials. To control impacts is important 
in ball games in sports and many processes in industrial plants. 
The impacts of such the
macroscopic materials cause complicated processes and eventually
they become inelastic.    

Studies of inelastic impacts are aimed to clarify the relation between 
the pre-collisional state and the post-collisional state.
Since there are huge number of 
degrees of freedom in macroscopic materials, it is
difficult to follow all the processes of energy transfers 
induced by the impact. In this situation, we need to introduce simple
quantities to characterize the inelastic impact of macroscopic materials. 

For this purpose,
the coefficient of restitution  $e$ 
is widely used\cite{newton,vincent,raman,goldsmith,stronge}, 
which is defined by
\begin{equation}\label{1}
{\bf v_c}^{'} \cdot {\bf n}=-e {\bf v_c} \cdot {\bf n},
\end{equation} 
where ${\bf v_c}$ and ${\bf v_c}^{'}$ are respectively 
the relative velocity 
at the contact point of two colliding materials 
before and after the collision, 
and ${\bf n}$ is the normal unit vector 
of the tangential plane of them (Fig.1). 
Although many text books of elementary physics state 
that $e$ is a material constant less than unity, 
it has been confirmed that $e$ decreases
as the impact 
velocity increases\cite{goldsmith,stronge,bridges,sonder,kuwabara}. 
For example, 
the dependence of $e$ for the low impact velocity are theoretically 
treated by the quasi-static theory
\cite{kuwabara,morgado,brilliantov96,schwager,ramirez}.

Rebound processes depend on the impact angle. Therefore, we also
introduce 
the coefficient of tangential restitution 
$\beta$ as
\begin{equation}\label{1-2}
{\bf v_c}^{'} \cdot {\bf t}=-\beta {\bf v_c} \cdot {\bf t},
\end{equation}
where ${\bf t}$ is the unit tangential vector (Fig.1). 
$\beta$ is a function of the incident angle 
$\gamma$ which is defined as $\gamma=\arctan(v_t/v_n)$
with $v_n={\bf v_c} \cdot {\bf n}$ and $v_t={\bf v_c} \cdot {\bf t}$.
It is believed
that possible values of $\beta$ lie
between -1 and 1\cite{walton_proc,walton_rheol,
labous,foerster,lorentz,gorham}. 
The relation between the impact speed ${\bf v}$
of the center of mass and ${\bf v}_c$ at the contact point is given by
\begin{equation}
{\bf v}_c={\bf v}-R {\Vec{\omega}}\times {\bf n}
\end{equation}
for hard spheres, where $R$ and $\Vec{\omega}$ are the radius of the
sphere and the angular velocity of the sphere, respectively. 
The phenomenological theories of the oblique impact 
have been developed\cite{walton_proc,walton_rheol,mbf76,stronge} 
and used 
in the explanation of results of experimental and numerical 
studies\cite{kuninaka_jpsj2003,mbf81,labous}.

\begin{figure}[htb]\label{t-fig3}
\includegraphics[width=8cm]{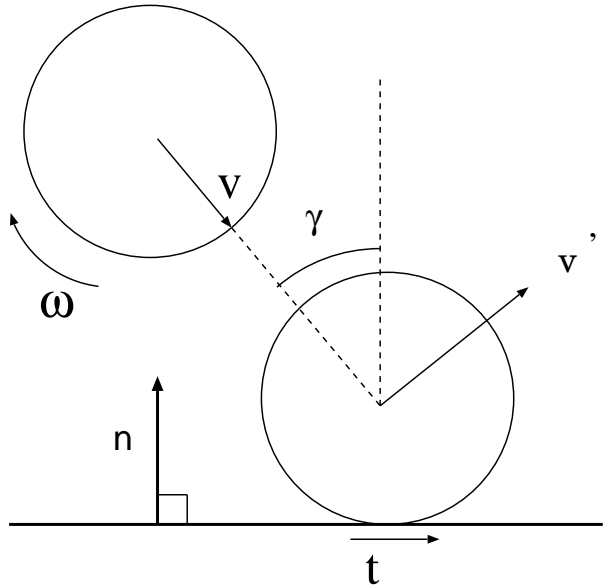}
\caption{
 A schematic picture of collision between a sphere and a flat wall.
}
\end{figure}

The impact process is important in granular physics. Although we believe
that the static interaction among contact grains can be described by the
contact mechanics of elastic materials, we little know the dynamical part
of contact mechanics. Thus, the distinct element method  which is the most
popular method for the simulation of grains contains many unknown
parameters.\cite{dem} Therefore, the theory of impact for
elastic  materials gives the basis of granular 
physics.\cite{jaeger96,granules,degennes}
 
While $e$ has been believed to be less than $1$ in most situations, 
we have recently recognized that $e$ can exceed $1$
in oblique impacts\cite{smith,calsamiglia,louge}.
In particular, Louge and Adams\cite{louge} reported
that $e$ increases as a linear function of 
the magnitude of $\tan\gamma$ in the oblique impact 
of a hard aluminum oxide sphere on a thick plate 
with the incident angle $\gamma$. 
In this case, Young's modulus of the wall is $100$ times smaller
than that of the sphere in the experiment. 
Thus, the physics of impact is one of interesting subjects in current
statistical mechanics.
 
The organization of this paper is as follows. 
In the next section, we explain the current understanding of normal
impacts including the quasi-static theory, the effect of radiation of
sounds, 
and the effect of the plastic deformation.
In section III, we summarize the standard treatments for oblique impacts 
which include Walton's argument\cite{walton_proc,walton_rheol} and 
the theory by Maw {\it et al.}\cite{mbf76} 
In section IV, we will introduce three typical approaches for simulation of impacts
of elastic materials. In section V, we briefly explain the recent analysis to explain the anomalous behavior of 
the restitution coefficient exceeds unity. Section VI is the short summary of this paper.  

\section{The current understanding on normal impact}

This section is devoted to summarize the current understanding on a normal 
head-on collision of elastic spheres or a collision between
a sphere and a wall. From the
quasi-static
theory\cite{kuwabara,brilliantov96,morgado,schwager,ramirez}, 
we believe that the restitution coefficient $e$
decays as $1-e\propto v^{1/5}$ with $v=v_n$ for small impact
velocity. The agreement between the theory and experimental 
results is fair.\cite{kuwabara,sonder,labous}
For high speed impacts, the plastic deformation of elastic
particles is dominant mechanism to determine the post-collisional
processes. Experimental results support the theoretical prediction
$e\propto v^{-1/4}$.\cite{wjohnson,johnson}  At present, we do not have any appropriate theory
 for the finite impact speed but below the threshold at which
plastic deformation takes place.


\subsection{Mechanism of inelastic collision}

Inelasticity arises from the transfer of
translational kinetic energy to internal degrees of freedom.
The dominant dissipative processes for low speed impacts are 
viscous effects of elastic materials, the heat conduction and 
the sound emission into or out of the elastic materials. 
When a local deformation with finite speed
takes place in a macroscopic material, the system is excited to a
nonequilibrium state and after that it is relaxed to an equilibrium
state. 
It is not easy to specify the microscopic origin of viscous effects or the
relaxation process,
 because such systems 
have numerous number of degree of freedom. However, at least, inelastic
scatterings of phonons and excitation-radiation processes in electronic
states are two major sources of the dissipations.

It should be noted that the coefficient of restitution of
one-dimensional rods is insensitive to the impact speed but depends on the
ratio of lengths for two colliding rods.\cite{goldsmith,giese,aspel}
On the other hand, the restitution coefficient strongly depends on
the impact speed and Poisson's ratio in higher dimensional impacts.
For example, the two-dimensional simulation by Gerl and Zippelius\cite{gerl}
shows that inelasticity increases as Poisson's ratio increases.

\subsection{Outline of theory of a quasi-static impact}

In this subsection, we present the outline of quasi-static theory for
a normal impact of elastic spheres.

The theory is based on the contact theory of elastic spheres developed by
Hertz.\cite{hertz} Hertzian contact theory predicts that the radius of contact $a$
and the elastic compress force
$F_{el}$ for two spheres with radii $R$ and $R'$ are respectively given by
\begin{equation}\label{hertz}
a= F_{el}^{1/3}\left(D \frac{RR'}{R+R'}\right)^{1/3}:\quad
F_{el}
=\frac{h^{3/2}}{D}\left(\frac{RR'}{R+R'}\right)^{1/2}
\end{equation}
where $h$ is the length of compression, and
$D=\frac{3}{4}(\frac{1-\nu}{E}+\frac{1-\nu'^2}{E'})$ with Young's
modulus $E, E'$ and Poisson's ratio $\nu,\nu'$ for two contact
materials.\cite{love,landau,hills,galin,johnson}

Let us consider a low speed impact of two spheres. In the limit of low 
speed, the nonequilibrium processes  may be suppressed, and the
collision can be treated as an elastic process.
The energy conservation can be read 
\begin{equation}\label{2-ene_cons}
m_{eff} \left(\frac{dh}{dt}\right)^2+\kappa h^{5/2}=m_{eff} v^2,
\quad \kappa=\frac{4}{5D}\displaystyle\sqrt{\frac{RR'}{R+R'}} ,
\end{equation}
where $m_{eff}=mm'/(m+m')$ is the reduced mass of two spheres with
masses $m$ and $m'$. The maximum deformation
is easily obtained as $h_0=(m_{eff}/\kappa)^{2/5}v^{4/5}$.
The contact time $t_c$ of the collision which is two times as large as
 the time needed to 
reach $h_0$ is  given by
\begin{equation}\label{tc}
t_c=2\left(\frac{m_{eff}^2}{\kappa^2 v}\right)^{1/5}
\int_0^1\frac{dx}{\sqrt{1-x^{5/2}}}=c_t 
\left(\frac{m_{eff}^2}{\kappa^2v}\right)^{1/5}
\end{equation}
where $c_t=4\sqrt{\pi}\Gamma(2/5)/5\Gamma(9/10)
\simeq 2.94$ with the Gamma function $\Gamma(x)$.


The above treatment predicts $e=1$, because the process does not contain
any dissipation. Kuwabara and Kono\cite{kuwabara} assume the existence
of Rayleigh's dissipation function for elastic solids, and write down
the dissipative force  as 
\begin{equation}\label{dissip-f}
F_f=-2\displaystyle\sqrt{\frac{RR'}{R+R'}}
\frac{\sqrt{h}}{\tilde D}\frac{dh}{dt}
\end{equation}
for two contacted spheres. Here, $\tilde D$
corresponds to $D$ for elastic contact, which can be represented by
viscous parameters of Rayleigh's dissipation function.
The magnitude of the viscous parameter $\tilde D$ can be measured from
an experiment of sound attenuation.
Adding this viscous term, 
the dimensionless form of 
equation of motion becomes
\begin{equation}\label{kk-9}
\ddot x+\eta \sqrt{x} \dot x+\frac{5}{4}x^{3/2}=0;
\quad \eta\equiv \frac{5}{2}\tilde \kappa 
\left(\frac{v}{\kappa^3m_{eff}^2}\right)^{1/5}
\end{equation}
where 
$\tilde\kappa=(4/5\tilde D)\sqrt{RR'/(R+R')}$.  
Here, we nondimensionalize the variables in terms of 
$h=h_0x$, $t=(h_0/v)\tau$, and thus $\dot x=dx/d\tau$.
Thus, the problem is reduced to obtaining $e=dx/d\tau$ at
$\tau_c=t_c(v/h_0)$ under the initial conditions $x=0$ and $\dot x=1$.
From (\ref{kk-9}), it is easy to obtain
\begin{equation}
e^2-1=-2\eta\int_0^{\tau_c}\sqrt{x}{\dot x}^2d\tau.
\end{equation}
Since the exact evaluation of the integral is impossible, we may replace
$x$ in the above equation by the solution of elastic equation. Using
this approximation we obtain
\begin{equation}
e\simeq 1-1.009\eta =1-1.009 \times \frac{5}{2}\tilde 
\kappa\left(\frac{v}{\kappa^3{m_{eff}}^2}\right)^{1/5} ,
\end{equation}
where 
the numerical constant comes from $(4/5)B(3/5,3/2)\simeq 1.009$ with the 
beta function $B(x,y)$. Thus, the coefficient of restitution for the low
speed impact is believed to decay $1-e\propto v^{1/5}$.
The result can be obtained in different contexts.\cite{morgado,brilliantov96}

Here, we briefly summarize the two-dimensional counterpart of normal impacts.
For impacts of an elastic disk on a rigid wall, we do not have reliable argument.
The total force of elastic force and dissipative force may be given by
\begin{equation}
F_{tot}\simeq -\frac{\pi E_*h}{\ln (4R/h)}-\tau_0\frac{\pi E_*\dot h}{\ln (4R/h)} ,
\end{equation}
where $E_{*}=E/(1-\nu^2)$ and
$\tau_0$ represents the time scale for the dissipation of the small
deformation. 
Replacing the logarithmic term as a constant
correction, the equation for elastic motion can be solved. 
Thus, we may evaluate the contact time $t_c$ as
\begin{equation}\label{t_c}
t_c\simeq \frac{\pi R}{c}\displaystyle\sqrt{\ln\frac{4c}{v}}
\end{equation}
where $c=\sqrt{E_*/\rho}$ and $\rho$ are  the compressive sound velocity and the density,
respectively.\cite{gerl,ces}  
From the comparison of eq.(\ref{t_c}) with a two-dimensional simulation,
we have confirmed that the above estimation 
is quantitatively correct.\cite{ces} 
Here, we adopt a bold approximation: 
 $h_{max}\simeq v R\sqrt{(\rho/E_*)\ln (4R/h)}\sim v R/c$.
Including the dissipative force  this approximation gives
\begin{equation}
e\simeq 1-\frac{2\pi E_*}{\rho R c\sqrt{\ln (4c/v)}}.
\end{equation}
The correctness of this expression has not been confirmed.

\subsection{The effect of radiation of elastic waves}

So far, we do not have any theoretical argument to estimate the
restitution coefficient as a result of radiation of elastic waves. On
the other hand, Miller and Pursey\cite{miller54,miller55} 
calculated the averaged power radiated 
per unit area by the normal oscillating contact of circular region on a
semi-infinite isotropic elastic material. Although the situation is a
little different each other, we may apply their calculation to estimate
the energy loss by radiation of elastic waves. 

Miller and Pursey\cite{miller54,miller55} obtained the powers 
radiated in the compressible ($W_c$) and shear waves ($W_s$) as
\begin{eqnarray}
W_c &=& \frac{a^4{k_1}^2t_c\zeta^4P_0^2}{4\rho }\int_0^{\pi/2}d\theta 
\{\Theta_1(\theta)\}^2\sin\theta \\
W_s &=& \frac{a^4{k_1}^2t_c\zeta^9P_0^2}{4\rho }\int_0^{\pi/2}d\theta 
|\Theta_2(\theta)|^2\sin\theta ,
\end{eqnarray}
where $\theta$ is the polar angle from the axis of symmetry, 
$P_0=F_{el}/(\pi a^2)$ is the average compressive force, and
\begin{eqnarray}
\Theta_1(\theta)&=&\frac{\cos\theta(\zeta^2-2\sin^2\theta)}{F_0(\sin\theta)} 
\quad
\Theta_2(\theta)=\frac{\sin2\theta\sqrt{\zeta^2\sin^2\theta-1}}{F_0(\zeta \sin\theta)} \\
F_0(x)&\equiv&(2x^2-\zeta^2)^2-4x^2\sqrt{(x^2-\zeta^2)(x^2-1)}, \\
k_1&=&\frac{\pi}{t_c}\sqrt{\frac{\rho(1+\nu)(1-2\nu)}{E(1-\nu)}},
\quad 
k_2=\frac{\pi}{t_c}\sqrt{\frac{2\rho(1+\nu)}{E}} \\
\zeta&=&k_2/k_1=\sqrt{\frac{2(1-\nu)}{1-2\nu}}.
\end{eqnarray}
Here we omit the power radiated in
terms of surface waves which plays an important role in the paper by
 Miller and Pursey\cite{miller54,miller55}, because they can be absorbed 
 as the numerical factor and the treatment for spherical surfaces is not 
 obvious.
The integrations of $W_c$ and $W_s$ are possible once we specify the
material. For example, in the case of the material with $\nu=1/4$, {\it i.e.}
$\zeta=\sqrt{3}$, the total amount of power is given by
\begin{equation}\label{miller18-19} 
W=W_c+W_s=1.579\frac{a^4{k_1}^2t_c\zeta^4P_0^2}{4\rho}.
\end{equation}

The energy loss during collision is thus given by
\begin{equation}
E_{loss}=2\int_0^{t_c/2}dt W\simeq 
\frac{5.337}{\rho t_c^2 E^{3/2}}\int_0^{t_c/2}dt F_{el}^2.
\end{equation}
Similarly the argument in the previous section, the integral in the
right hand side can be evaluated as $\int_0^{t_c/2}dt F_{el}^2=
5\sqrt{\pi}\Gamma(8/5)/(8\Gamma(21/10))\kappa^{2/5}m_{eff}^{8/5}v^{11/5}$
in the elastic limit.
Thus, we obtain
\begin{equation}
E_{loss}\simeq 0.5827 \frac{\kappa^{6/5} m_{eff}^{4/5}}{\rho E^{3/2}}v^{13/5}.
\end{equation}
From the relation $E_{loss}=(1-e^2)m_{eff}v^2/2\simeq m_{eff}v^2(1-e)$,
we finally reach the new relation:
\begin{equation}\label{new}
1-e\simeq 0.5827 \frac{\kappa^{6/5}}{\rho m_{eff}^{1/5} E^{3/2}}v^{1/5}, 
\end{equation}
for $\nu=1/4$. 
The discussion here is the rough evaluation of the restitution coefficient
based on the energy loss by radiation of elastic waves. Although 
the numerical constant in eq.(\ref{new}) is meaningless, we expect that 
the scaling relation can be used in realistic situations. It is
interesting that the restitution coefficient $e$ obeys $1-e\propto
v^{1/5}$ which is the essentially same as that  of the quasi-static 
theory.\cite{kuwabara,morgado,brilliantov96,schwager,ramirez}    

It should be noted that Hunter\cite{hunter} derived 
$1-e\propto v^{3/5}$ from the 
theory of Miller and Pursey.\cite{miller54,miller55} 
The difference comes from the followings.
Hunter\cite{hunter} assumes that the energy dissipation is obtained from 
\begin{equation}
W_H=\int_{0}^{t_c}dt \pi a^2 F_{el}(t)\frac{d\bar u(t)}{dt},
\end{equation}
where $\bar u(t)$ is the mean surface displacement. 
Although Hunter\cite{hunter}
adopts the result of Miller and Pursey\cite{miller54,miller55} for $\bar u$, this is only the
fraction of the radiation. The above treatment presented here is more
reliable.  

However, the analysis presented in this subsection is still
prematured. Insufficient parts of the analysis are as follows: (i) The
theory by Miller and Pursey\cite{miller54,miller55} assume the constant pressure in the contact
area, but Hertzian contact theory predicts the distribution of
pressure. (ii) Miller and Pursey\cite{miller54,miller55} 
discussed the radiation of elastic
wave for semi-infinitely large region, but the actual contacted spheres
have curvature and finite volume. Therefore, we may need more systematic 
treatment for the radiation of elastic waves.

\subsection{Impact with plastic deformation}

When the impact speed is large enough, the elastic description is no longer
valid but we have to consider the effects of plastic deformation. 
The restitution coefficient drastically decreases when the plastic
deformation occurs. Following the argument by Johnson\cite{johnson},
 we review the
argument of the restitution coefficient for impacts with plastic
deformations. In the argument in this subsection, we neglect numerical
factors. So the result is basically valid for collisions for two spheres 
with equal radius $R$ and the density.

Let us evaluate the
yield pressure. From Hertzian contact theory the maximum pressure of
compression exists at the center of contact and its expression is given by
\begin{equation}
 p_0=\frac{3F_{el}}{2\pi a^2}=\left(\frac{6F_{el}{E_*}^2}{\pi^3R^2}\right)^{1/3}.
\end{equation}
Thus, the compressive elastic force at the yield pressure $p_0=(p_0)_Y$
becomes 
 \begin{equation}
 F_{el}=\frac{\pi^3R^2}{6{E_*}^2}(p_0)_Y^3, 
\end{equation}
where $(p_0)_Y$ is about 1.6 $E$ at Mises' condition.\cite{johnson}

Hertzian contact theory gives $F_{el}\sim E_*R^{1/2}h^{3/2}$, while 
the maximum deformation is 
\begin{equation}
 h_*\sim \left(\frac{m v^2}{R^{1/2}E_*}\right)^{2/5}
\end{equation}
from $m v^2\sim F_{el}h\sim 
E_* R^{1/2}h^{5/2}$, where $m$ is the mass of each sphere.
 Substituting this $h_*$ into Hertzian contact theory 
with setting $F_{el}=F_Y$, we balance the force with  
$F_Y\sim \frac{R^2}{{E_*}^2}(p_0)_Y^3$. Then we obtain
\begin{equation}
(p_0)_Y \sim \left(\frac{E_*}{R^{3/4}}\right)^{4/5}(mv^2)^{1/5},
\end{equation}
which is the condition for yield stress.

Let us discuss the effect of plastic deformation to the restitution
coefficient. From Hertzian contact theory 
there is a relation between the compression and the radius of the contact 
$h \sim a^2/R$. Therefore the deformation $h_*$ for the plastic
deformation becomes $h_*\sim F_*/(E_*a_*)$ because of 
$F_*\sim E_* \sqrt{Rh} h$, 
 where $a_*$ and $F_*$ are
respectively the radius of contact area and the compressive force for
the plastic deformation. Thus, the internal energy stored during the
deformation may be evaluated as 
$W'\sim F_*h_*\sim
{F_*}^2/(a_*E_*)$. On the other hand, we have 
$F_*=\pi a_*^2p_d$ with the contact pressure $p_d$.
Since the stored energy is released as the kinetic energy in impact
processes, the kinetic energy for rebound is given by
$K_r\sim W'\sim a_*^3p_d^2/E_*$.

On the other hand, 
we can estimate the work needed for compression as
$W=(1/2)mv^2 \sim
     \int_{0}^{h_{*}}dhF_{el} \sim 
     \int_{0}^{a_{*}}da F_*a/R$. Assuming that $F_{el}=\pi a^2 p_d$ is
     kept during the impact, we obtain
\begin{equation}\label{11.42}
 W=\int_0^{a_*}\frac{\pi p_d a^3da}{R}\sim \frac{ p_d {a_*}^4}{R}\sim mv^2.
\end{equation}
Thus, we reach
\begin{equation}\label{e2}
e^2=\frac{{v'}^2}{v^2}\sim \frac{p_d R}{E_*a_*} .
\end{equation}
Equation (\ref{e2}) can be rewritten as
\begin{equation}
e^2\sim \frac{p_d}{E_*}\left(\frac{p_dR^3}{m v^2}\right)^{1/4}
\end{equation}
and thus, we finally obtain
\begin{equation}\label{1/4law}
e\propto v^{-1/4}.
\end{equation}
This expression recovers experimental results.


\section{Current Understanding of Oblique Impact}

\subsection{Walton's argument}

To characterize the oblique collision,
 Walton introduced three parameters:\cite{walton_proc,walton_rheol}
 the coefficient of normal restitution $e$, 
 the coefficient of Coulomb's friction $\mu$,
 and the maximum value of the coefficient of tangential
 restitution $\beta_0$.
 The expression is given by
\begin{equation}\label{walton32}  
\beta \simeq
\left\{
 \begin{array}{ll}
  -1-\mu (1+e) \cot\gamma \left(1+\frac{mR^2}{I}\right)
& \quad (\gamma \ge \gamma_{0})\\
   \beta_{0} & \quad  (\gamma \le \gamma_{0}),
 \end{array}
\right.
\end{equation}
where $\gamma_0$ is the critical angle, and $m$, $R$, and $I$ are
 mass, radius and moment of inertia of spheres
 respectively.\cite{walton_proc,walton_rheol}
 Experiments have supported that his characterization
 adequately capture the essence of binary collision of spheres or
 collision of a sphere on a flat plate\cite{labous,foerster,lorentz,gorham}.

The derivation of the first equation of Walton's expression is simple.
When there is a slip, the friction coefficient satisfies
\begin{equation}\label{j6} 
|{\bf n}\times {\bf J}|=\mu ({\bf n}\cdot{\bf J}) ,
\end{equation}
where ${\bf J}=m({\bf v}'-{\bf v})$ is the impulse.
Let us write ${\bf J}$ as the following form:
\begin{equation}\label{j7}
{\bf J}=A({\bf n}\cdot {\bf v}_c){\bf n}+B[{\bf v}_c-
({\bf n}\cdot {\bf v}_c){\bf n}] ,
\end{equation}
where $A$ and $B$ are constants to be determined. From the definition of 
$e$ and $\beta$ with the aid of 
$I({\Vec \omega}'-\Vec{\omega})=-R({\bf n}\times {\bf J})$
we write
\begin{equation}\label{j8}
m ({\bf v}_c'-{\bf v}_c)=A({\bf n}\cdot {\bf v}_c){\bf n}+(1+\frac{mR^2}{I})B[{\bf v}_c-({\bf n}\cdot {\bf v}_c){\bf n}].
\end{equation}
Thus, from the projection to the normal direction we obtain $A=-m(1+e)$ 
and $B=(1+mR^2/I)m(1+\beta)$. On the other hand, through
the relation 
$|{\bf n}\times {\bf J}|=B|{\bf n}\times {\bf v}_c|=\mu ({\bf n}\cdot {\bf 
J})=\mu A({\bf n}\cdot {\bf v}_c)$,
we can rewrite ${\bf J}$ as
\begin{equation}\label{j9}
{\bf J}=-m(1+e)({\bf n}\cdot{\bf v}_c){\bf n}+\mu m(1+e)\cot\gamma [{\bf v}_c
-({\bf n}\cdot {\bf v}_c){\bf n}].
\end{equation}
Thus, we obtain the first expression of eq.(\ref{walton32}).

\begin{figure}[htbp]
\begin{center}
\includegraphics[width=0.5\textwidth]{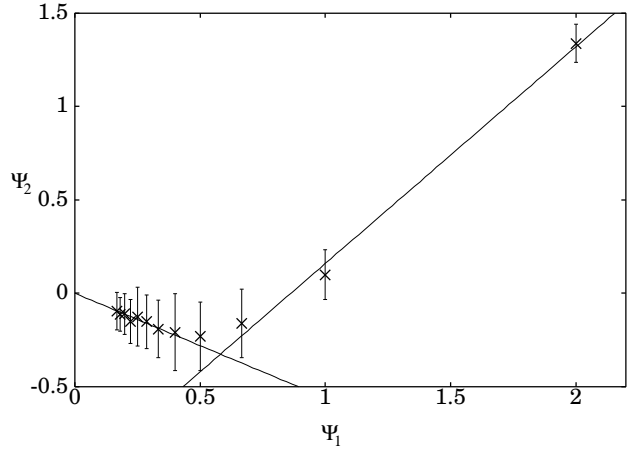}
\end{center}
\caption{Comparison between Walton's argument and our numerical
 result. Here $\Psi_1=\tan\gamma$ and $\Psi_2=-{v_t}'/v_n$.
Parameters for the simulation are the same as those 
in ref.\cite{kuninaka_jpsj2003}.}
\label{walton}
\end{figure}
In spite of its simple form and its simple derivation, Walton's
expression is useful to characterize oblique impacts. However, we do not 
know how to determine $\beta_0$ and $\gamma_0$. In addition, it cannot be
applied to impacts for very small $\gamma$ in which $\beta$ cannot be a
constant, because it should not be contradict with the normal impact at
$\gamma=0$. The discontinuity of differentiation of $\beta$ at 
$\gamma=\gamma_0$ is also unnatural. To know the details of physics of
the oblique impacts we should introduce alternative theory.

\subsection{Theory of Maw {\it et al}.}

More systematic treatment for oblique impacts is developed by Maw {\it
et al.}\cite{mbf76,stronge,kuninaka_jpsj2003} The complete explanation of their theory is long  because of its
complicated structure.  Here, we only summarize the result of
the theory. 

They assume that 
the stiffness of normal compliance for restitution changes from $k$ to
$k/e^2$ where $k$ is the normal stiffness for compression. 
They do not
discuss the origin of the normal restitution coefficient $e$ which is
assumed to be constant and its
effect for the stiffness explicitly. 

They also indicate that there are three region depending on the incident angle $\gamma$.
According to their theory\cite{mbf76}, 
$\beta$ for the impact of a circular disk on a wall can be represented
by\cite{kuninaka_jpsj2003}

\begin{enumerate}[(i)]
 \item $1/\mu \sigma^2 < \cot \gamma$ with $\sigma=\sqrt{(2-\nu)/(2(1-\nu))}$:\hspace{3mm}
\begin{equation}\label{regime1}
\beta=-\cos \omega t_1 - \mu \frac{\beta_x}{\beta_z} e
  \left[1 + \cos \left( \frac{\Omega t_1}{e}+
                   \frac{\pi}{2}(1-e^{-1})\right) \right] \cot \gamma,
\end{equation}
 \item $\beta_x/\beta_z \mu(1+e) < \cot \gamma < 1/\mu \sigma^2$:\hspace{3mm}
\begin{eqnarray}\label{regime2}
  \beta &= &-\cos \omega (t_3-t_2)-\mu \frac{\beta_x}{\beta_z}
  [\cos \omega (t_3-t_2)
  -\cos \Omega t_2 \cos \omega (t_3-t_2) \nonumber\\
 & &+ \frac{\Omega}{\omega} \sin \Omega t_2
  \sin \omega (t_3-t_2)
  + e + \cos \Omega t_3
  ] \cot \gamma,
 \end{eqnarray}
 \item $ \cot \gamma < \beta_x/\beta_z \mu(1+e)$:\hspace{3mm}
\begin{equation}\label{regime3}
\beta=-1 + \mu \frac{\beta_x}{\beta_z}
  \left(1 + e \right) \cot \gamma,
\end{equation}
\end{enumerate}
where $\mu$ is the coefficient of friction,  
 $\beta_x=3$ and $\beta_z=1$ are constants for the impact of the disk on the infinite wall.  
 $\Omega$ and $\omega$ are respectively $\pi/2t_0$
 and $(\pi/2\sigma t_0)\sqrt{\beta_x/\beta_z}$,
 where $t_0$ is the time for compression.
The time $t_1$ 
 is the transition time from initial stick motion to slip motion
which is determined by
\begin{equation}
\frac{|F_x(t_1)|}{\mu F_z(t_1)}=
\left\{ 
 \begin{array}{ll}
   \displaystyle\frac{1}{\sigma^2} \frac{v_x(0)}{\mu v_z(0)}
   \frac{\Omega}{\omega} \frac{\sin \omega t_1}{\sin \Omega t_1}=1
   & 0 \leq t_1 < t_0 \\
   \displaystyle\frac{1}{\sigma^2} \frac{v_x(0)}{\mu v_z(0)}
   \frac{\Omega}{\omega}
   \frac{\sin \omega t_1}{\sin (\frac{\Omega t_1}{e}+
   \frac{\pi}{2}(1-e^{-1}))}=1
   & t_0 \leq t_1 < t_f\label{a10} ,
 \end{array}
\right.
\end{equation}
where $v_x(t)$ and $v_z(t)$ are the tangential and the normal relative
velocities at the contact point, respectively. 
While $t_2$ can be determined by (\ref{a20}) :
\begin{equation}
\begin{array}{lr}
\Omega t_2 = \arccos \left(
 \displaystyle\frac{v_x(0)/\mu v_z(0)-\beta_x/\beta_z}{\sigma^2-\beta_x/\beta_z}
\right)&{\rm for }{~}\displaystyle\frac{v_x(0)}{v_z(0)} \leq \mu \displaystyle\frac{\beta_x}{\beta_z}\\
\displaystyle\frac{\Omega t_2}{e} = -\displaystyle\frac{\pi}{2}(1-e^{-1})
+\arccos \left(
 \frac{v_x(0)/\mu v_z(0)-\beta_x/\beta_z}{\sigma^2 e^{-1} - e \beta_x/\beta_z}
\right)&{\rm for } {~}\displaystyle\frac{v_x(0)}{v_z(0)} > \mu \displaystyle\frac{\beta_x}{\beta_z}
\label{a20}
\end{array}
\end{equation}
which is the time to start sticking.
The time $t_3$ determined by solving eqs.(\ref{a21}) numerically
 is the transition time from stick motion to slip motion:
\begin{equation}
\left| \frac{\Omega u_x(t_2)}{\mu v_z(0)}\cos \omega(t_3-t_2)
-\frac{\Omega v_x(t_2)}{\omega \mu v_z(0)}\sin \omega(t_3-t_2) \right|
=\sigma^2 \sin\left[\frac{\Omega t_3}{e}+\frac{\pi}{2}(1-e^{-1})\right]
\label{a21},
\end{equation}
where $u_x(t_2)$ is the tangential deformation at time $t_2$.
 By calculating $\beta$ at each value of $\cot\gamma$ and interpolating
 them with cubic spline interpolation method,
 we can draw the theoretical curve. 
\begin{figure}[htbp]
\begin{center}
\includegraphics[width=0.5\textwidth]{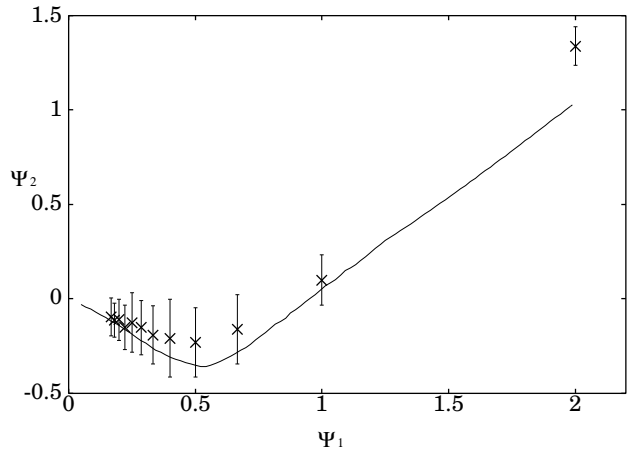}
\end{center}
\caption{Comparison between Maw's theory and our numerical result. The
 situation is the same as that in ref.\cite{kuninaka_jpsj2003}.}
\label{maw}
\end{figure}
Figure \ref{maw} shows comparison of the theory with our numerical simulation, 
in which the agreement is acceptable.

As is shown, the theory by Maw {\it et al.}\cite{mbf76,stronge}
 captures the physics of impact processes. For large $\gamma$, {\it i.e.} region 
(iii), the disk slips on the wall without any rotation or sticking. For
 intermediate $\gamma$, that is, the region (ii), the disk
slips at first and stick at $t=t_2$ and slips again at $t=t_3$. 
For small $\gamma$ in the region (i), the disk sticks initially and begins to 
slip at $t=t_1$. In this sense, their theory improves some defects of 
Walton's argument.\cite{walton_proc,walton_rheol} 
However, their theory includes some 
other defects: The final expression for $\beta$ is complicated and is required for numerical calculation. It still contains
undetermined parameters $e$ and $\mu$ which are assumed to be constants.
As is shown, $e$ and $\mu$ strongly
depend on the impact speed and 
the incident angle.\cite{kuwabara,morgado,brilliantov96,sonder,labous,vincent,schwager,ramirez,gerl,ces,gorham,louge,kuninaka_prl}
 Thus, we cannot justify  their assumption.  

\section{Numerical modeling}

In general,
the success of theoretical approach is limited for nonlinear problems
because of difficulties
of analysis. On the other hand, numerical
approach has become standard as computers become popular.
The advantage of this approach is obvious. (i) We can investigate
collision processes under idealistic situations. (ii) It is easy to
control situations to investigate the properties of impact
processes. (iii) It is possible to analyze nonlinear problems. 
However, when we restrict our interest in numerical studies of impacts,
we still do not have any standard technique and the status of such the
studies is prematured.

One of typical approaches for engineers is to use FEM (Finite Element
Method).\cite{smith,lim,lim99,minamoto,wu} There are some standard packages for simulation of 
impact processes. For example, Lim and Stronge\cite{lim,lim99} carried out
two-dimensional simulation of a transverse collision of a cylinder
against an elastoplastic half space based on DYNA2D. A
three-dimensional FEM also exists.\cite{minamoto} 
All of them reproduce
experimental results. Since FEM is originally proposed for a solver for static
problems, they can recover the Hertzian contact theory
for the static elastic problem. 
No viscous term, however,
 is included within FEM. To obtain inelastic impacts, FEM usually introduces
elastic-plastic deformation or fully plastic deformation. Therefore, the 
simulation based on FEM sometimes predicts $e\propto v^{-1/4}$ without
quasi-static region.\cite{lim} 
We also note that there are many input parameters for FEM. For example, it includes at least two yield stresses for 
the transition from the elastic region to the elastic-plastic region, and  the transition from the elastic-plastic region to
the fully plastic region. It also contains at least two friction coefficient as a constant. However, as will be shown, 
the friction coefficient cannot be regarded as a constant. 

Another approach is proposed by Gerl and Zippelius\cite{gerl,ces} which is
based on the mode analysis of elastic materials.  They assume that
an isolated disk is in an eigenstate of isothermal vibration of elastic
waves. When the disk contacts a wall, the transitions among eigenstates
is induced by nonlinear effects of interaction between the disk and the
wall. 
 
The last approach is based on the equation of motion for mass points
connected by springs.\cite{ces,kuninaka_jpsj2003} 
This approach is intuitive and flexible. For
example, the other approaches may not be applied to impacts with rough surfaces
without essential change of algorithms, but this approach can include
the roughness at surface easily. The roughness at
surface plays crucial roles for oblique impacts. 
In fact, the coefficient of tangential restitution $\beta$ becomes -1
without the roughness.\cite{kuninaka_jpsj2003}
The disadvantage of
this approach is that the code is not fast, and the result strongly
depends on lattice structures.

The last two approaches are adequate for the high speed impact without
plastic deformation. At present, there is no three dimensional
simulations, and two-dimensional simulations for normal impacts
 may not agree with those
for quasi-static theory. One of defects for these approaches is that the
problem is not relaxed to a static state without introduction of local
dissipation. For later discussion, we focus on the last approach to simulate the oblique impact of a disk on a flat wall 
to clarify the mechanism of anomalous behavior of the restitution coefficient $e$.

\section{Simulation of oblique impact}

\begin{figure}[htbp]
\begin{center}
\includegraphics[width=0.6\textwidth]{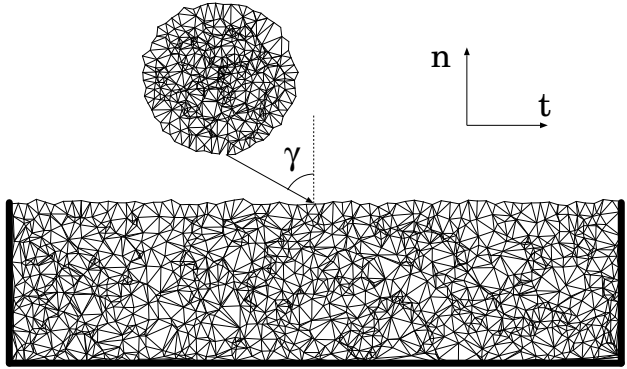}
\end{center}
\caption{The elastic disk and wall consisted of random lattice.}
\label{fig1}
\end{figure}
\subsection{Our model}

Let us introduce our numerical model\cite{kuninaka_jpsj2003,kuninaka_prl}
which is a typical example of the last approach in the previous section. 
The discussion in this section is based on our recent paper.\cite{kuninaka_prl}

Our numerical model consists of an elastic disk and an elastic wall 
(Fig. \ref{fig1}). 
The width and the height of the wall are $8 R$ and $2 R$, respectively, 
where $R$ is the radius of the disk. 
We adopt the fixed boundary condition for the both side ends and the bottom 
of the wall. 
To make each of them, at first, we place mass points at random 
in a circle and a rectangle with the same density, respectively. 
For the disk, we place $800$ particles at random in a circle 
with the radius $R$ 
while for the wall, similarly, we place $4000$ particles 
at random in a rectangle.

After that, we connect all mass points with nonlinear springs 
for each of them using the Delaunay triangulation
algorithm\cite{delaunay}.
The spring interaction between connected mass points 
is given by
\begin{equation}\label{interaction}
 V^{(i)}(x)=
\frac{1}{2} k^{(i)}_{a} x^{2}+\frac{1}{4} k^{(i)}_{b} x^{4},
\hspace{3mm}i=d\text{(disk)},w\text{(wall)},
\end{equation} 
where $x$ is a stretch from the natural length of spring, 
and $k^{(i)}_{a}$ and $k^{(i)}_{b}$ are the spring constants 
for the disk($i$=d) and the wall($i$=w).

In most of our simulations, 
we adopt $k^{(d)}_a = 1.0 \times m_0 c^2/R^2$for the disk 
while $k^{(w)}_a = k^{(d)}_a /100$ for the wall, 
where $m_0$ and $c$ are the mass of each mass point and 
the one-dimensional velocity of sound, respectively. 
In this model, the wall is much softer than the disk 
as in ref.\cite{louge}. 
We adopt $k^{(i)}_b = k^{(i)}_a \times 10^{-3}/R^2$ 
for each of them.
We do not introduce any dissipative mechanism in this model. 
The interaction between the disk and the wall during a collision 
is given by 
${\bf F}(l)=\xi V_0\exp(\xi l){\bf n}^{s}$, 
where $\xi$ is $300/R$, $V_0$ is $\xi m_0c^2R/2$, 
$l$ is the distance between each surface particle of the disk 
and the surface spring of the wall,  
and ${\bf n}^{s}$ is the normal unit vector to the spring. 

In this model, the roughness of the surfaces is important 
to make the disk rotate after collisions\cite{kuninaka_jpsj2003}. 
To make roughness, the normal random numbers with its average is zero 
and its standard deviation is $\delta = 3 \times 10^{-2} R$
are used for the surface particles of the disk and the wall. 
All the data in this paper are obtained from the average  
of 100 samples in random numbers. 

Poisson's ratio $\nu$ and Young's modulus $E$ of this model 
can be evaluated from the strains of the band of random lattice 
in vertical and horizontal directions to the applied force. 
We obtain Poisson's ratio $\nu$ and Young's modulus $E$  
as $\nu=(7.50 \pm 0.11) \times 10^{-2}$ 
and $E=(9.54 \pm 0.231)\times 10^{3}m_0c^{2}/R^{2}$, 
respectively.\cite{kuninaka_jpsj2003} 

In our simulation, we define the incident angle $\gamma$ 
by the angle between the normal vector of the wall and the initial 
velocity vector of the disk(see Fig. \ref{fig1}). 
We fix the initial colliding speed of the disk 
as $|{\bf v}(0)|=0.1c$ 
to control the normal and tangential components of 
the initial colliding velocity as $v_t(0)=|{\bf v}(0)|\sin\gamma$ and 
$v_n(0)=|{\bf v}(0)|\cos\gamma$, respectively. 
We use the fourth order symplectic numerical method 
for the numerical scheme of integration 
with the time step $\Delta t = 10^{-3}R/c$.

\begin{figure}[htbp]
\begin{center}
\includegraphics[width=0.5\textwidth]{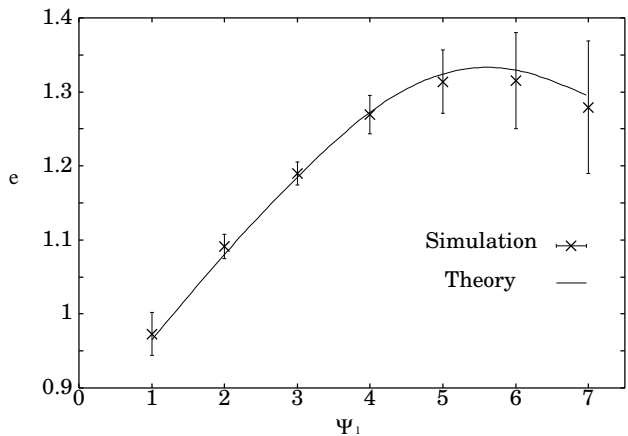}
\end{center}
\caption{Numerical and theoretical results of the relation between
 $\Psi_1$ and $e$.}
\label{fig2}
\end{figure}
\subsection{Results}
Figure \ref{fig2} is the normal restitution coefficient $e$ against 
$\Psi_1=\tan\gamma$ for the impact of the hard disk on the soft wall. 
The cross points are the average and the error bars are  
the standard deviation of 100 samples for each incident angle. 
This result shows that $e$ increases as $\Psi_1$ increases 
to exceed unity, and has a peak around $\Psi_1 = 6.0$.
This behavior is contrast to that in the experiment 
by Louge and Adams\cite{louge}.

Let us clarify the mechanism of our results. 
Louge and Adams\cite{louge} suggest the anomalous behavior of $e\ge 1$ 
can be understood by the local deformation 
on the surface of the wall during an impact. 
They attribute their results to the rotation of normal unit vector 
of the wall surface by an angle $\alpha$ 
and derive the corrected expression for $e$. 
Thus, we determine the quantity of $\alpha$ at each incident angle 
from the theory of elasticity and 
calculate corrected $e$.

\begin{figure}[htbp]
\begin{center}
\includegraphics[width=0.5\textwidth]{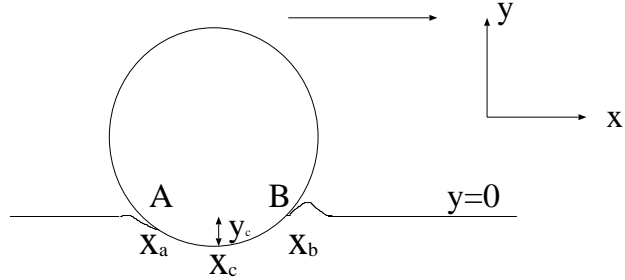}
\end{center}
\caption{The schematic figure of a hard disk sliding on a soft wall. 
$x$ coordinates of both ends of the contact area AB are $x=x_a$ and 
$x=x_b$.}
\label{fig3}
\end{figure}

Figure \ref{fig3} is the schematic figure of a hard disk moving 
from left to right on a wall,  
where the length of the contact area is $l=|x_b-x_a|$. 
From the theory of elasticity\cite{hills,galin}, 
this ratio can be estimated as 
\begin{equation}\label{ratio1}
 \frac{x_c-x_a}{l} = 1-\theta \hspace{5mm} \text{with} \hspace{5mm}
 \theta = \frac{1}{\pi}\arctan \frac{1-2\nu}{\mu (2-2\nu)},
\end{equation}
where $\nu$ is Poisson's ratio of the wall and $\mu$ is the coefficient
of friction. 
Here  
$\tan\alpha\equiv \frac{f(x_b)-f(x_a)}{l}$ with the deformed shape of
the wall $f(x)$ approximated by a parabolic function near $x_c$ is reduced to 
\begin{equation}\label{tanalpha2}
\tan \alpha = \frac{1-2\theta}{2-2\theta} \frac{|x_c-x_a|}{R}.
\end{equation}
In eq.(\ref{tanalpha2}), 
$|x_c-x_a|$ can be evaluated by the simulation data. 
From our simulation, the maximum value of $y_c$ is about $0.17R$ 
at $\Psi_1=1.0$. 
Assuming the disk is pressed in the normal direction, 
we can estimate the contact area as about $1.1R$ 
which is the maximum value.
Thus, we adopt its half value, $0.55R$, as $|x_c-x_a|$. 

\begin{figure}[htbp]
\begin{center}
\includegraphics[width=0.5\textwidth]{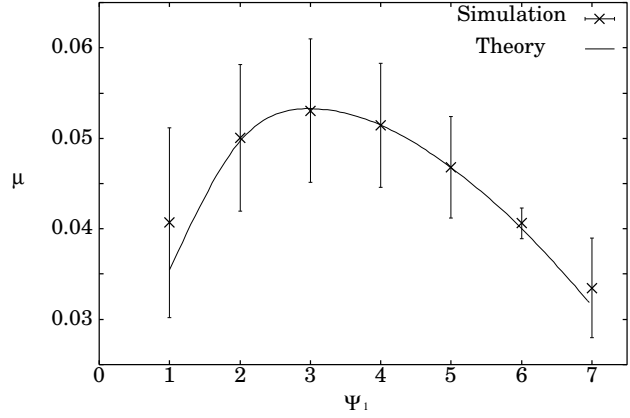}
\end{center}
\caption{Numerical and theoretical results of the relation between
 $\Psi_1$ and $\mu$.}
\label{fig4}
\end{figure}

The cross points in Fig.\ref{fig4} is $\mu$ calculated 
from eq.(\ref{j6}) and our simulation data against each $\Psi_1$. 
Figure \ref{fig4} shows $\mu$ has a peak around $\Psi_1=3.0$. 
Substituting this result to eqs.(\ref{ratio1}) 
and (\ref{tanalpha2}), 
we obtain the relation between $\Psi_1$ and $\tan\alpha$.

The restitution coefficient $e$ can be obtained as  a function of $\tan\alpha$ by regarding the impact as 
that on a tilted surface with the angle $\alpha$. 
Skipping the derivation, we can write the result as\cite{kuninaka_prl}
\begin{equation}\label{CORmod2}
e=\frac{e_{\alpha}+\Psi_2^{\alpha} \tan\alpha}{1-\Psi_1^{\alpha} \tan\alpha},
\end{equation}
where $e_{\alpha}$ is the restitution coefficient defined through
\begin{equation}
{\bf v_c}'\cdot {\bf n}_{\alpha}=-e_{\alpha}({\bf v_c}\cdot {\bf n}_{\alpha}),
\end{equation}
where ${\bf n}_{\alpha}$ is the unit normal vector of the tilted slope
connecting $A$ and $B$ in Fig.6.
Here  
$\Psi_1^{\alpha}$ 
and $\Psi_2^{\alpha}$ are respectively given by 
\begin{equation}\label{psi1}
\Psi_{1}^{\alpha}=\frac{\Psi_1-\tan\alpha}{1+\Psi_1 \tan\alpha} . 
\end{equation}
and
\begin{equation}\label{from_wal}
\Psi_2^{\alpha}=\Psi_1^{\alpha}-3(1+e_{\alpha})\mu_{\alpha}
\end{equation}
in the two-dimensional situation\cite{walton_proc,walton_rheol}.
$\mu_{\alpha}$ in eq.(\ref{from_wal}) is given by
\begin{equation}
\mu_{\alpha}=\frac{\mu+\tan\alpha}{1-\mu \tan\alpha}.
\end{equation}
To draw the solid line in Fig.\ref{fig2}, at first, 
we calculate $\tan\alpha$ and 
$\mu$ by eqs.(\ref{tanalpha2}) and (\ref{j6}) respectively for each $\Psi_1$. 
After that, we calculate $\Psi_1^{\alpha}$ and $\Psi_2^{\alpha}$ 
by eqs.(\ref{psi1}) and (\ref{from_wal}), and obtain $e$ by substituting 
them into eq.(\ref{CORmod2}) for each $\Psi_1$. 
Although $e_{\alpha}$ is assumed to be a constant in ref.\cite{kuninaka_prl},
here we evaluate $e_{\alpha}$ from our simulation which has the weak
dependence of $\gamma$.
The solid line of Fig.\ref{fig2} is eq.(\ref{CORmod2}). 
All points are interpolated with cubic spline interpolation method 
to draw the theoretical curve. 
Such the theoretical description of $e$ 
is consistent with our numerical result. 
As can be seen, the restitution coefficient 
$e$ depends on the relation between 
$\mu$ and $\Psi_1$.  

It should be noted that the behavior of $\mu$ as a function of $\gamma$ can be understood by using a simple 
phenomenological argument. The solid curve in Fig.\ref{fig4} is the
fitting curve obtained from the theory. If we omit the collapse of the
roughness by the impact for large $\gamma$ close to $\pi/2$,
both $\mu$ and $e$ increase with increasing $\gamma$. This suggests that 
the discrepancy between our result in Fig.5 and the result by Louge and
Adams\cite{louge} is originated from the difference of impact
speed. Namely, our impact speed much larger than the experimental one. 
Although we skip the details of derivation, it may be clear that our
argument captures the essence of physics of the oblique impact.

\subsection{Discussion}

At first, let us discuss the origin of the relation 
between $e$ and $\Psi_1$. 
As was indicated, 
the local deformation of the wall's surface is essential 
for $e$ to exceed unity. 
We have also carried out the simulation when 
$k^{(w)}_a = 10 \times k^{(d)}_a$, which means 
the wall is harder than the disk. 
In this case, $e$ takes almost constant value to exceed unity suddenly 
around $\Psi_1=4.5$. This tendency resembles the experiment by Calsamiglia 
{\it et al.}\cite{calsamiglia}. 
This behavior can be understood as that the disk is scattered by
  hard nails distributed on the surface of the wall.
\begin{figure}[htbp]
\begin{center}
\includegraphics[width=0.5\textwidth]{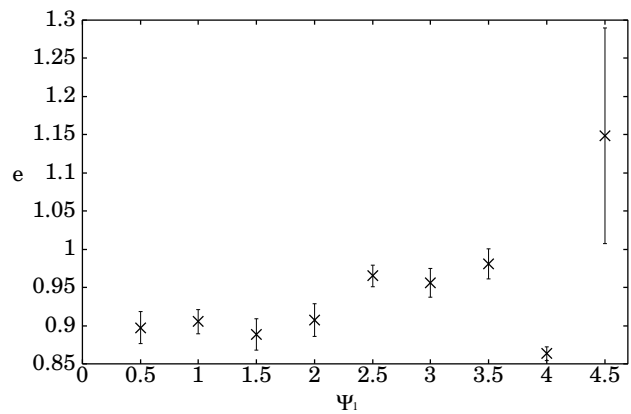}
\end{center}
\caption{Numerical results of $e$ for the impact of a soft disk on a
 hard wall. It should be noted that $\Psi_1=4.5$ is almost the same as
 the largest $\gamma$ in the experiment of ref.\cite{calsamiglia}.}
\label{hard}
\end{figure}

Thus, 
the wall should be much softer than 
the disk to get smooth increases of $e$ as increasing $\gamma$. 
In addition, 
it is important to fix the initial kinetic energy of the disk. 
We have confirmed so far 
that $e$ cannot exceed unity when we change $\gamma$
 with fixed
$v_n$\cite{kuninaka_jpsj2003}. 

Second, the initial velocity of the disk and the local deformation 
of the wall are so large that the local dissipation in springs and 
the gravity have not affected our numerical results. 
In addition,  we have carried out the other simulation 
with a disk of $400$ particles 
and a wall of $2000$ particles to investigate the effect of the model
size. 
Although there is a slight difference between the results, 
the data can be reproduced quite well by our phenomenological theory.  

Third, the local deformation of the wall also affects 
the relation between $\mu$ and $\Psi_1$. 
In early studies, 
it has been shown that $\mu$ depends on the impact velocity
\cite{gorham,louge}. 
In our simulation, the magnitude of $\alpha$ has a peak 
around $\Psi_1=3.0$. 
This behavior is interpreted as 
that the local deformation collapses for large $v_t$.
The decrease of $\mu$ and the friction force cause the decrease of $e$.

In final, we adopt the static theory of elasticity to explain our
numerical results for the discussion here. 
However, it is important to solve the time-dependent equation 
of the deformation of the wall surface 
to analyze the dynamics of impact phenomena. 
The dynamical analysis is our future task.

In summary of this section, we have carried out the two-dimensional simulation 
of the oblique impact of an elastic disk on an elastic wall. 
We have found that the restitution coefficient $e$ can exceed unity in the oblique impact, 
which is attributed to the local deformation of the wall. 
The relation between $\mu$ and $\Psi_1$ is also 
related to the local deformation and can be explained by a simple theory.

\section{Summary}

We review the current understanding on inelastic impacts of elastic materials. 
We explain the standard theories for the normal impact and for oblique impacts. We also introduce some typical
approaches for numerical simulation for impact problems. 
Although this problem is fundamental and
familiar in elementary mechanics, the theoretical treatment is prematured. In recent finding of the anomalous 
behavior of the restitution coefficient exceeds unity is an example 
to have potential to be developed as a subject of physics.
Unfortunately, most of the existing theories are based on engineering idea and they are not so simple and beautiful. 
In some cases, the assumption of the theory may not be valid. For physicists, in other words, there will be a lot of
room for improvement of the theory of impact of elastic materials. 

\begin{acknowledgments}
We would like to thank M. Y. Louge for fruitful discussion and
 N. Mitarai for her critical reading of this manuscript.
We also appreciate Sanjay Puri for giving us the opportunity to write this paper.  
Parts of numerical computation in this work were carried out 
at Yukawa Institute Computer Facility. 
This study is partially supported by the Grant-in-Aid of 
Ministry of Education, Science and Culture, Japan (Grant No. 15540393).
\end{acknowledgments}
\bibliographystyle{apsrev}
\bibliography{impact}

\end{document}